\documentclass[apj]{emulateapj}

\def\gtorder{\mathrel{\raise.3ex\hbox{$>$}\mkern-14mu
             \lower0.6ex\hbox{$\sim$}}}
\def\ltorder{\mathrel{\raise.3ex\hbox{$<$}\mkern-14mu
             \lower0.6ex\hbox{$\sim$}}}

\usepackage{colordvi}

\slugcomment{Draft of \today}

\shorttitle{SN\,2010jl}

\shortauthors{Ofek et al.}

\begin{document}

\title{A six year image-subtraction light curve of SN\,2010jl}
\author{E.~O.~Ofek\altaffilmark{1},
B.~Zackay\altaffilmark{2},
A.~Gal-Yam\altaffilmark{1},
J.~Sollerman\altaffilmark{3},
C.~Fransson\altaffilmark{3},
C.~Fremling\altaffilmark{4},
S.~R.~Kulkarni\altaffilmark{4},
P.~E.~Nugent\altaffilmark{5,6},
O.~Yaron\altaffilmark{1},
M.~M.~Kasliwal\altaffilmark{4},
F.~Masci\altaffilmark{7},
R.~Laher\altaffilmark{7}}

\altaffiltext{1}{Benoziyo Center for Astrophysics and the Helen Kimmel center for planetary science, Weizmann Institute of Science, 76100 Rehovot, Israel.}
\altaffiltext{2}{School of Natural Sciences, Institute for Advanced Study, 1 Einstein Drive, Princeton, New Jersey 08540, USA}
\altaffiltext{3}{Department of Astronomy, The Oskar Klein Centre, Stockholm University, AlbaNova University Centre, SE-106 91 Stockholm, Sweden}
\altaffiltext{4}{Cahill Center for Astronomy and Astrophysics, California Institute of Technology, Pasadena, CA 91125, USA}
\altaffiltext{5}{Computational Cosmology Center, Lawrence Berkeley National Laboratory, 1 Cyclotron Road, Berkeley, CA 94720, USA}
\altaffiltext{6}{Department of Astronomy, University of California, Berkeley, CA 94720-3411, USA}
\altaffiltext{7}{Infrared Processing and Analysis Center, California Institute of Technology, Pasadena, CA 91125, USA}

\begin{abstract}

SN\,2010jl was a luminous Type IIn supernova (SN), detected
in radio, optical, X-ray and hard X-rays.
Here we report on its six year $R$- and $g$-band light curves
obtained using the Palomar Transient Factory.
The light curve was generated using a pipeline based on
the proper image subtraction method and we discuss the algorithm
performances.
As noted before, the $R$-band light curve,
up to about 300\,days after maximum light
is well described by a power-law decline
with a power-law index of $\alpha\approx-0.5$.
Between day 300 and day 2300 after maximum light, it is
consistent with a power-law decline, with a power-law index of about
$\alpha\approx-3.4$.
The longevity of the light curve suggests that the
massive circum-stellar material around the progenitor
was ejected on time scales of at least tens of
years prior to the progenitor explosion.

\end{abstract}


\section{Introduction}
\label{sec:Introduction}

The emission from some supernovae (SNe) is powered by the conversion of the
SN-ejecta kinetic energy into visible light over time scales
of months to years (orders of magnitude
shorter than in SN remnants). This mechanism is
responsible for powering the light curves of at least
some Type IIn SNe (e.g., 
Schlegel et al. 1990;
Chevalier \& Fransson 1994;
Chugai \&  Danziger 1994;
Ofek et al. 2013a;
see Filippenko 1997; Gal-Yam 2017 for the definition of Type IIn SNe).

SN\,2010jl (PTF\,10aaxf) is a Type IIn SN discovered
by Newton \& Puckett (2010).
The SN took place in a star-forming galaxy (UGC~5189A)
at a distance of about 50\,Mpc.
Reaching a visual magnitude of approximately 13, this SN was observed
across the electromagnetic spectrum
(e.g., Patat et al. 2011;
Smith et al. 2011;
Stoll et al. 2011;
Chandra et al. 2012, 2015;
Zhang et al. 2012;
Moriya et al. 2013;
Ofek et al. 2013b, 2014a;
Fransson et al. 2014;
Aartsen et al. 2015;
Ackermann et al. 2015;
Jencson et al. 2016).
The SN radiated $\gtorder6\times10^{50}$\,erg
(Fransson et al. 2014), likely from the conversion
of the kinetic energy in the ejecta to visible light
via interaction of the ejecta with circumstellar material (CSM)
around the SN progenitor.
Visible light and X-ray observations contain evidence that
the CSM is likely very massive
(e.g., Zhang et al. 2012;
Chandra et al. 2012, 2015;
Fransson et al. 2014; Ofek et al. 2014a)
with order-of-magnitude mass-estimates in the range of 5--15\,M$_{\odot}$.
{\it NuSTAR} hard X-ray observations of this system
can be used to estimate
the shock velocity, which was found to be consistent with the
estimates based on the visible light data (Ofek et al. 2014a),
and with the spectroscopic estimates (Borish et al. 2015).

Recently, Fox et al. (2017) presented late time {\it HST} observations
that constrain the progenitor luminosity.
Dwek et al. (2017) use this, as well as {\it Spitzer} observations,
to constrain the dust mass around the progenitor.
They argue that if the progenitor is assumed to be similar to $\eta$~Car,
then about 4\,magnitudes of extinction are required,
which suggests $\gtorder10^{-3}$\,M$_{\odot}$ of dust around
the progenitor, prior to its explosion.
Gall et al. (2014) argue, based on an extinction derived from
the supernova line ratios, 
that at late times (868\,days)
about $2\times10^{-3}$\,M$_{\odot}$ were formed around SN\,2010jl
(assuming dust is mainly composed of Carbon, but could be up to an order of magnitude larger for silicates).
This estimate assumes that the conditions
at the line-formation site
do~not vary with time.
Sarangi et al. (2018) estimated a dust mass of order
$10^{-2}$\,M$_{\odot}$ (at an age of few hundreds days after explosion).
Sarangi et al. (2018) estimated a dust mass of
$2\times10^{-3}$--$10^{-2}$\,M$_{\odot}$,
depending on dust composition, at day 844 after the explosion.

Given the long timescales over which SN\,2010jl is visible,
this object presents a unique opportunity to study
the physics of collisionless shocks propagating within massive
CSM (e.g., Katz et al. 2011; Murase et al. 2011, 2014).
Here we present a six year
light curve of SN\,2010jl.
We measure the SN light curve using the 
proper image subtraction algorithm of Zackay, Ofek, \& Gal-Yam (2016; ZOGY),
and we evaluate the performance of this method in a complex environment.

In \S\ref{sec:Obs} we present the SN observations.
In \S\ref{sec:pipe} we describe our image subtraction photometry pipeline,
while in \S\ref{sec:test} we discuss its performances,
and we conclude in \S\ref{sec:disc}.

\section{Observations and light curve}
\label{sec:Obs}

The Palomar Transient Factory (PTF)
and its continuation project the intermediate PTF (iPTF;
Law et al. 2009; Rau et al. 2009),
using the 48-inch Oschin Schmidt Telescope,
observed the field of SN\,2010jl over 600 times.
The data reduction is described in Laher et al. (2014),
while the photometric system is discussed in Ofek et al. (2012a).

The SN is located on top of a bright star-forming region ($r\approx15.5$\,mag).
The bright host galaxy makes its necessary to use image-subtraction
based photometry.
We constructed image-subtraction-based light curves
in the Mould $R$ and $g$ bands, based on all the images in which the transient
location is more than 100 pixels from a CCD edge.
In total we used 485 Mould $R$-band images and 185 $g$-band
images\footnote{The images were taken at two PTF fields and CCDs.
PTF field 3159 and CCDID 6, and PTF field 100072 and CCDID 10.}.
The analysis was performed using tools available in the
MATLAB astronomy and astrophysics
toolbox\footnote{https://webhome.weizmann.ac.il/home/eofek/matlab/}
(Ofek 2014).

For the reference image, we selected all the images taken prior to 2010 May 23.
This amounts to 27 $R$-band and 12 $g$-band images.
The image-subtraction photometry pipeline is briefly described in
\S\ref{sec:pipe}. The photometric calibration was done against
the Pan-STARRS-1 catalog (PS1; Chambers et al. 2016).

\begin{figure*}
\centerline{\includegraphics[width=16cm]{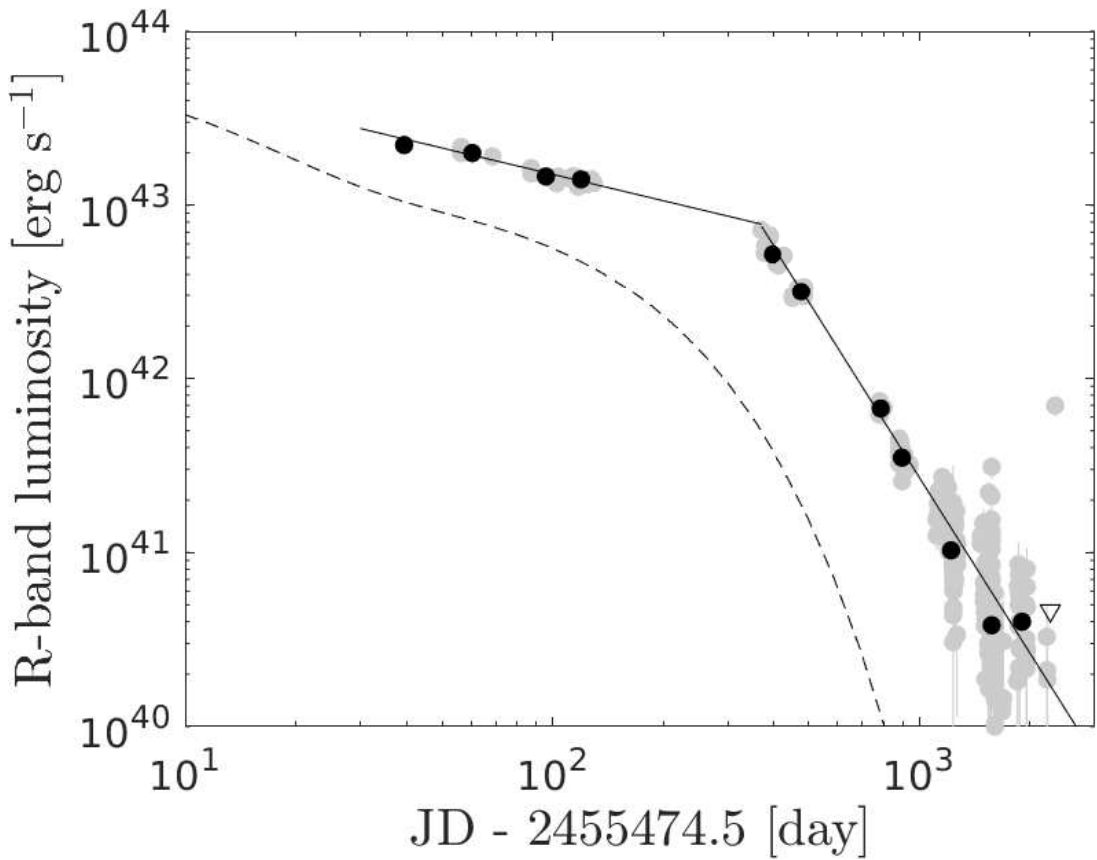}}
\caption{The PTF $R$-band light curve of SN\,2010jl (gray circles).
The black circles are binned photometry (including negative flux measurements),
while the empty-black triangle represents a binned 3-$\sigma$
upper limit on the luminosity.
The luminosity light curve is corrected for Galactic extinction.
The dashed line shows the expected radiated bolometric luminosity from
one solar mass of radioactive Nickel 56.
\label{fig:SN2010jl_R}}
\end{figure*}

\begin{figure}
\centerline{\includegraphics[width=8cm]{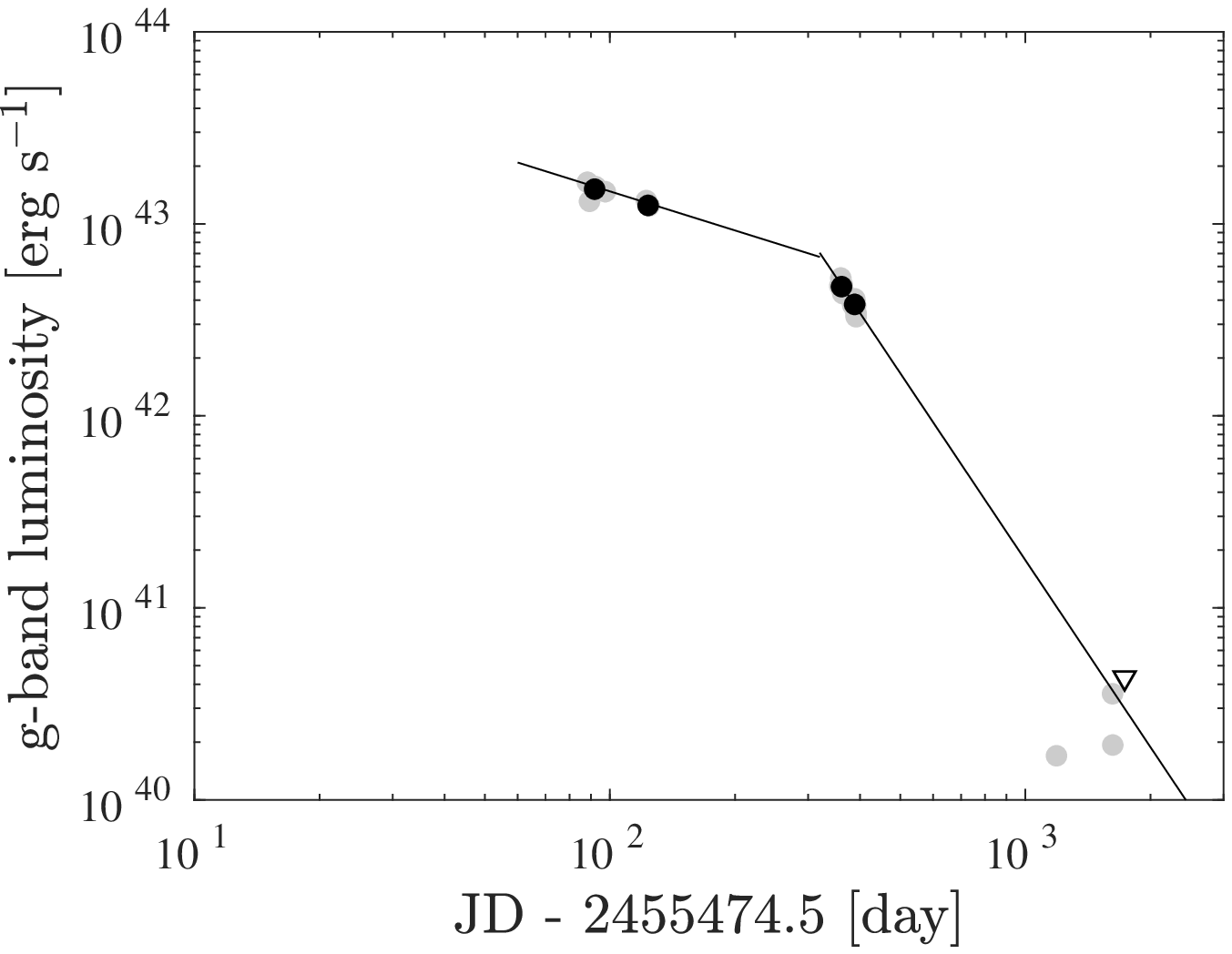}}
\caption{Like Figure~\ref{fig:SN2010jl_R}, but for the $g$-band.
The individual measurements after 1000\,days are consistent
with no detection.
\label{fig:SN2010jl_g}}
\end{figure}

The image-subtraction-based measurements are listed in Table~\ref{tab:PTFphot},
and the entire $R$-band light curve is presented in
Figure~\ref{fig:SN2010jl_R},
while the $g$-band light curve is shown in Figure~\ref{fig:SN2010jl_g}.
The photometry in the plots are corrected for Milky-Way extinction
($E_{\rm B-V}=0.027$\,mag; Schlegel et al. 1998; Cardelli et al. 1989).
\begin{deluxetable*}{lllllll}
\tablecolumns{7}
\tablewidth{0pt}
\tablecaption{PTF Photometric measurements of SN\,2010jl}
\tablehead{
\colhead{Band}          &
\colhead{JD-JD$_{0}$}    &
\colhead{Mag}           &
\colhead{Mag Err.}      &
\colhead{Counts}        &
\colhead{Counts error}  &
\colhead{$\chi^{2}_{epochal}$} \\
\colhead{}            &
\colhead{(day)}       &
\colhead{(mag)}       &
\colhead{(mag)}       &
\colhead{(count)}     &
\colhead{(count)}     &
\colhead{}
}
\startdata
$g$  &   $-563.8546$ &   NaN  &  NaN & $ -7.56\times10^{3}$ & $ 6.73\times10^{3}$ &   991.0\\
$g$  &   $-563.6819$ &   NaN  &  NaN & $  8.24\times10^{1}$ & $ 3.61\times10^{3}$ &   674.6\\
$g$  &   $-562.7827$ &   NaN  &  NaN & $ -3.43\times10^{3}$ & $ 4.39\times10^{3}$ &  1141.3\\
$g$  &   $-562.6970$ &   NaN  &  NaN & $  8.08\times10^{2}$ & $ 7.73\times10^{3}$ &  1242.3\\
$g$  &   $-557.8225$ &   NaN  &  NaN & $ -4.10\times10^{3}$ & $ 8.99\times10^{3}$ &   814.9
\enddata
\tablecomments{Image-subtraction-based
photometry of SN\,2010jl.
JD$_{0}=2455474.5$, corresponding to about 20\,days prior to
$I$-band maximum light (Stoll et al. 2011).
The PS1-based photometric zero points, not corrected for color term,
are $30.213$ and $32.443$, for $g$ and $R$ band, respectively.
The magnitudes are shown in Luptitude units (Lupton, Gunn, \& Szalay 1999)
in order to deal with negative fluxes.
Note that there is a small color term, between the PS1 and PTF magnitudes
(see Ofek et al. 2012a for PTF filters transmission).
Photometry is done in the PTF native photometric system
(i.e., the color term was fitted, and the color of the
SN was set to zero).
This table is published in its entirety in the electronic edition. A portion of the full table is shown here for
guidance regarding its form and content.}
\label{tab:PTFphot}
\end{deluxetable*}

We fit a broken power-law to the light curves.
Each power law fit is of the form
\begin{equation}
L = L_{0}(t-t_{0})^{-\alpha},
\label{eq:PL}
\end{equation}
where $L$ is the luminosity, $t$ is the time in Julian Days (JD), $t_{0}$
was set to JD$=2455474.5$ (see justification in Ofek et al. 2014a),
and $\alpha$ is the power-law index.
The best-fit power-law indices are listed in Table~\ref{tab:PLfit}.

\begin{deluxetable}{lll}
\tablecolumns{3}
\tablewidth{0pt}
\tablecaption{SN\,2010jl power-law fit}
\tablehead{
\colhead{Band}          &
\colhead{Time range}    &
\colhead{$\alpha$}      \\
\colhead{}            &
\colhead{(day)}       &
\colhead{}            
}
\startdata
$R$  & 0--360     & $-0.50\pm0.0.04$  \\
$R$  & 360--1000  & $-3.36\pm0.04$  \\
$g$  & 0--360     & $-0.68\pm0.1$ \\
$g$  & 360--1000  & $-3.2\pm0.3$ 
\enddata
\tablecomments{Power-law index fits to SN\,2010jl light curve
in specific time ranges and bands, relative to $JD_{0}=2455474.5$.
To estimate the uncertainty on the fitted power law,
the individual photometric errors were renormalized such that
the $\chi^{2}$ per degree of freedom will be one.
Since in $g$ band the SN is detected only for about three years after
maximum light, we fit power laws only in the first 1000\,days.}
\label{tab:PLfit}
\end{deluxetable}

\section{Image-subtraction pipeline}
\label{sec:pipe}

Our pipeline\footnote{Implemented in {\tt ImUtil.pipe.imsub\_lightcurve}.}
is based on the image subtraction algorithm of
Zackay, Ofek, \& Gal-Yam (2017; ZOGY), and it contains the following steps:
Image cutouts of about $1000\times1000$\,pix, containing the
requested target, are read into memory.
The images are converted to units of electrons,
by multiplying the images by their gain.
The background is estimated both locally on scales of $64\times64$~arcsec,
and globally.
We then apply {\tt mextractor} to estimate the PSF for each image,
extract the sources and measure their position, shapes,
aperture photometry and PSF photometry (Ofek et al. in prep.).

We solve the astrometry of the images using {\tt astrometry.m} (Ofek 2018),
in respect to GAIA-DR2 reference stars (GAIA Collaboration et al. 2018),
and use SWarp (Bertin 2010) to interpolate the images to the same grid.
The reference image is photometrically calibrated using
either SDSS (Ahn et al. 2014),
PS-1 (Flewelling et al. 2016) or APASS (Henden et al. 2015) catalogs,
and we also fit for relative zero points between the images
(e.g., Ofek et al. 2011 Appendix A).
Next, we populate the mask image, associated with each science image,
with bits indicating saturated pixels, flat-field
holes\footnote{Flat-field holes are negative features in the science image generated by leftover sources in the flat image.},
and cosmic rays.
We read the observing date and exposure time
from the images header and calculate the mid-exposure Julian Day (JD).
A reference image is constructed either using proper coaddition
or simple weighted coaddition (Zackay \& Ofek 2017a,b).
The astrometric noise relative to the reference image is estimated,
and we perform the image subtraction,
and read the PSF photometry at the target position using Equation~41 in ZOGY.
Our pipeline also provide meta data information (see \S\ref{sec:test}).

One difference from the ZOGY algorithm
is that we propagate the astrometric errors
not only into the score\footnote{Denoted $S$ in ZOGY.} image, but also
into the relative uncertainty in the photometry.
Any astrometric errors, due to e.g., scintilation noise,
will effect the subtraction and hence the photometry.
The error in the PSF photometry is linear with the astrometric uncertainty.
Therefore,
the effect of such errors on the photometry
can be estimated from the image gradients.
Specifically, the additional fractional
error in the flux due to the astrometric uncertainty is
\begin{equation}
\sigma_{\rm astrom} = \frac{\sqrt{(\sigma_{x}\nabla_{x}{S_{N}})^{2} + (\sigma_{y}\nabla_{y}{S_{N}})^{2}}}{F_{S}}.
\label{eq:sigma_astrom}
\end{equation}
Here $\sigma_{x}$ and $\sigma_{y}$ are the astrometric uncertainties
in the $x$ and $y$ positions, respectively, measured in pixels;
$\nabla_{x}$ and $\nabla_{y}$ are the gradients in the $x$ and $y$ directions,
respectively; $S_{N}$ is given by Equation~31 in ZOGY.
and $F_{S}$ is the flux normalization of the image subtraction
statistics (Equation~42 in ZOGY).

\section{Testing the light curve}
\label{sec:test}

Here we present some of the sanity tests we performed on the light curve.
We demonstrate that such tests are useful for identifying problems
and should be used routinely.

\begin{figure}
\centerline{\includegraphics[width=8cm]{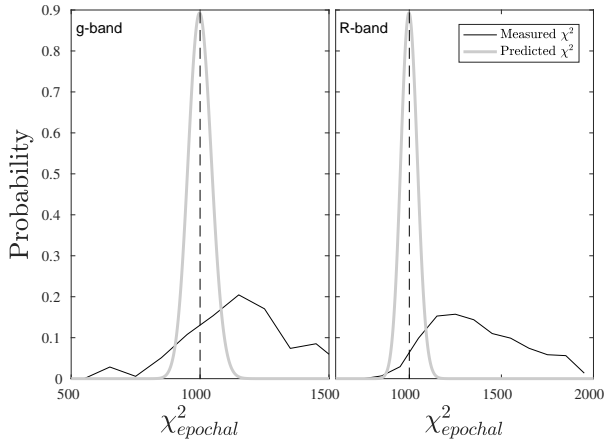}}
\caption{The epochal $\chi^{2}$ for $g$ band (left) and $R$ band (right).
\label{fig:SN2010jl_epChi2}}
\end{figure}

For all the images taken in each band,
we also generated light curves for 1000 random image positions.
We use these light curves to calculate two properties.
The first is the {\it epochal} $\chi^{2}$ presented in
Figure~\ref{fig:SN2010jl_epChi2}.
This is the $\chi^{2}$ over all 1000 random positions in one epoch,
where the errors in the $\chi^{2}$
are obtained using Equation 41 in Zackay et al. (2016).
If the fluctuations in the background of the subtracted image
in each epoch are represented by the error estimate,
then this epochal $\chi^{2}$ should be of the order of the
number of degrees of freedom (about 1000).
We note that we do~not expect that the epochal $\chi^{2}$ will be distributed
exactly like a $\chi^{2}$ distribution with the relevant number
of degrees of freedom.
This is because, in our ZOGY implementation we used a global background
variance value, while in practice the variance is slightly position
dependent.
This allows us to identify epochs in which the photometry
is highly uncertain.
Furthermore, we use it to correct the photometric errors
by a multiplicative factor of
\begin{equation}
\max{[1, \sqrt{\chi^{2}_{\rm epochal}/{\rm dof}}]}.
\label{eq:corr}
\end{equation}

\begin{figure}
\centerline{\includegraphics[width=8cm]{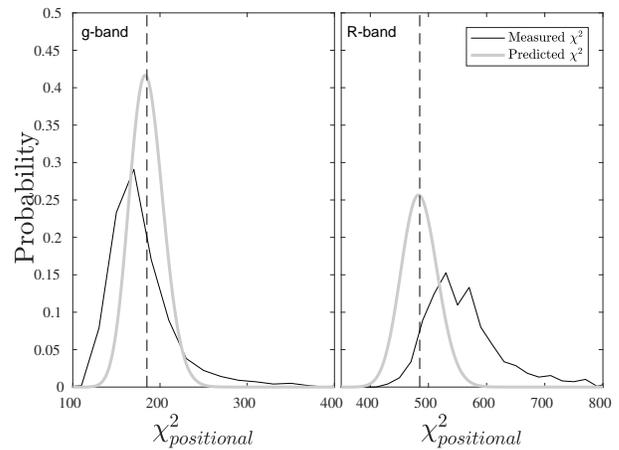}}
\caption{The positional $\chi^{2}$ for $g$ band (left) and $R$ band (right).
\label{fig:SN2010jl_posChi2}}
\end{figure}

The second property is the {\it positional} $\chi^{2}$ (Figure~\ref{fig:SN2010jl_posChi2}).
This is the $\chi^{2}$ in each random position over all
epochs. This is useful in order to identify issues
related to background estimation.

Figures~\ref{fig:SN2010jl_epChi2}-\ref{fig:SN2010jl_posChi2}
identify some epochs with bad subtractions,
and epochs in which our errors were under-estimated.
We corrected such errors using Equation~\ref{eq:corr}.
The underestimation of the errors is likely due to
errors in the flux matching process, background subtraction, PSF estimation,
and source noise due to the host galaxy flux.
We note that we propagated the photometric errors due to
the astrometric uncertainty using Equation~\ref{eq:sigma_astrom}.

Another interesting test is to correlate the residuals from the power-law fit,
when the SN is bright at the first 360 days, with various parameters.
We attempted to correlate the flux residuals with parameters
like the airmass,
flux matching ($\beta$), the mean level of the subtraction
image, the derived flux normalization
($F_{\rm S}$; Equation 42 in ZOGY),
and the epochal $\chi^{2}$.
Spearman rank correlations of the residuals from the best-fit
power law with these parameters were consistent with zero,
with false-alarm probability smaller than 10\% in all cases.

\section{Discussion}
\label{sec:disc}

In \S\ref{sec:BPL} we discuss the implication of the six-year light curve,
the pre-explosion observations are presented in \S\ref{sec:pre},
and the implications for the pre-explosion mass loss are discussed
in \S\ref{sec:loss}.

\subsection{The six-year light curve of SN\,2010jl}
\label{sec:BPL}

As noted by Ofek et al. (2014a),
the SN light curve is consistent with a broken power-law light curve.
%
Figures~\ref{fig:SN2010jl_R}--\ref{fig:SN2010jl_g} show
the $R$- and $g$-band light curves, respectively, of SN\,2010jl.
The gray lines represent the best-fit broken power laws to the $R$-band data.
The most important feature is that six years after maximum light,
the SN is still detected in the $R$ band and that the
late-time ($t-JD_{0}>1000$\,days)
light curve follows the power law fitted in the 360--1000-day range.

In the case of SN\,2010jl, a shock breakout likely occurred within
the CSM -- a so-called wind shock breakout
(e.g., Ofek et al. 2010; Chevalier \& Irwin 2011).
The hydrodynamics of ejecta with a power-law velocity distribution
moving into a CSM with a density profile that follows
another power-law distribution
is described by an analytical self-similar solution
(Chevalier 1982).
This hydrodynamical solution dictates the rate of kinetic energy
conversion into thermal energy and radiation
(e.g., Fransson 1984; Chugai \&  Danziger 1994; Svirski et al. 2012;
Moriya et al. 2013; Ofek et al. 2014a), which is yet another power law.
The light curve of SN\,2010jl at early times (about one year prior to maximum light)
is consistent with a power-law decay with a power-law index of
$\alpha\approx-0.5$.
Assuming spherical symmetry, power-law density distributions of the
CSM and ejecta, negligible bolometric correction\footnote{This is consistent with the roughly constant effective temperature reported in Ofek et al. (2014a).},
and using the Chevalier (1982) self-similar solution,
the observed power-law index suggest a CSM density profile of
$\approx r^{-2.2}$ to $r^{-2.3}$ for radiative/convective
stars (see e.g., Ofek et al. 2014a).

We note that the exact value of the power law slope
depends on $t_{0}$ and any unknown bolometric corrections
(see Ofek et al. 2014a for the dependence of the power-law
on $t_{0}$ and bolometric correction).
There are several possible explanations for the
broken power-law light curve
(e.g., geometry, van Marle 2010; and variations in the density
profile, Chandra et al. 2015).
However, it is not clear to us what is the correct
explanation for the discontinuity in the optical light curve of SN\,2010jl.

\subsection{Pre-explosion variability}
\label{sec:pre}

In recent years a large number of precursors --
outbursts prior to the SN explosion,
mainly prior to Type IIn SNe, were reported
(Foley et al. 2007; Pastorello et al. 2007;
Mauerhan et al. 2013; Pastorello et al. 2013;
Corsi et al. 2014;
Fraser et al. 2013;
Ofek et al. 2013b, 2014b, 2016;
Strotjohann et al. 2016;
Nyholm et al. 2017;
Arcavi et al. 2017).
Furthermore, Ofek et al. (2014b) showed that these precursors
are common in the final years prior to an explosion of a Type IIn SN.
SN\,2010jl was included in the sample of Ofek et al. (2014b),
and no precursor was found.
The amount of pre-explosion data we have for this SN
is small compared with other SNe in the Ofek et al. (2014b) sample.
Furthermore, it is possible that the amplitude
of any variability will be attenuated by contribution
from the underlying bright star-forming region or dust.

\begin{figure}
\centerline{\includegraphics[width=8cm]{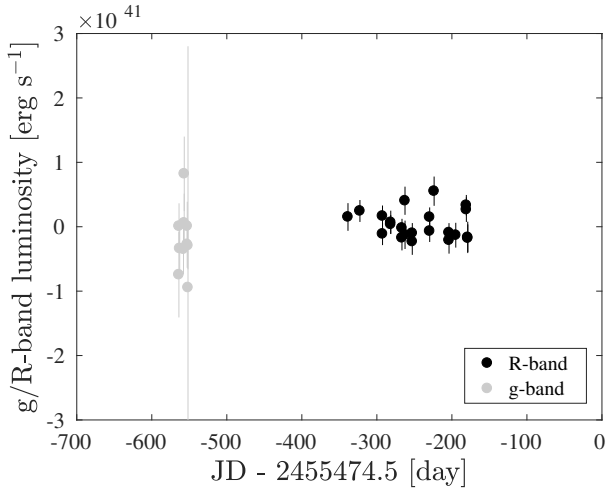}}
\caption{The pre-explosion $R$ (black circles) and $g$-band (gray circles)
light curves of SN\,2010jl.
\label{fig:SN2010jl_preexpLC}}
\end{figure}

Figure~\ref{fig:SN2010jl_preexpLC} shows the pre-explosion light curve at the
SN location. Since these observations were also used as a reference image,
all we can say is that the progenitor did~not show short-term variability
(i.e., smaller than a few weeks).
We set a 5-$\sigma$ upper limit of absolute magnitude
of $-13.8$ and $-13.9$ in $g$ and $R$ bands, respectively,
for any short term variability during these observations.
These absolute magnitudes are corrected only for Galactic extinction.

\subsection{Implications for pre-explosion mass-loss}
\label{sec:loss}

The late-time light curve of SN\,2010jl is still remarkably bright
in comparison with any reasonable contribution from $^{56}$Ni
(see Fig.~\ref{fig:SN2010jl_R}).
Since, at late times ($\approx400$\,days),
the shock ejecta velocity is of the order of 3000--5000\,km\,s$^{-1}$
(Ofek et al. 2014a), this implies that there is still a considerable
density of CSM at distances of $\sim2\times10^{16}$\,cm from
the SN.
Assuming a CSM velocity of 100\,km\,s$^{-1}$ (Fransson et al. 2014),
we conclude that the CSM
was ejected of the order of
\begin{equation}
\sim60 \Big(\frac{v_{\rm CSM}}{100\,{\rm km\,s}^{-1}}\Big)^{-1}\,{\rm yr},
\label{eq:Tcsm}
\end{equation}
prior to the SN explosion, where $v_{\rm CSM}$ is the CSM ejection velocity.
We note, for comparison, that the age of the $\eta$~Car
Homunculus Nebula is estimated to be about 1800\,yr
(Morse et al. 2001; Smith et al. 2017),
and that Sarangi et al. (2018) argued for the existstence
of a cavity in SN\,2010jl's CSM.

\acknowledgments

This paper is based on observations obtained with the
Samuel Oschin Telescope as part of the Palomar Transient Factory
project, a scientific collaboration between the
California Institute of Technology,
Columbia University,
Las Cumbres Observatory,
Oskar Klein Centre,
the Lawrence Berkeley National Laboratory,
the National Energy Research Scientific Computing Center,
the University of Oxford, and the Weizmann Institute of Science.
E.O.O. is grateful for support by
grants from the 
Willner Family Leadership Institute
Ilan Gluzman (Secaucus NJ),
Israel Science Foundation,
Minerva,
BSF, BSF-transformative,
Weizmann-UK,
and the I-Core program by the Israeli Committee for Planning
and Budgeting and the Israel Science Foundation (ISF).
B.Z. is grateful for receiving the support of the Infosys membership fund.

\end{document}